\documentclass[sigconf]{acmart}
\acmConference[EASE 2025]{The 29th International Conference on Evaluation and Assessment in Software Engineering}{17–20 June, 2025}{Istanbul, Türkiye}
\usepackage{packages}
\AtBeginDocument{%
  }

\setcopyright{acmlicensed}
\copyrightyear{2018}
\acmYear{2018}
\acmDOI{XXXXXXX.XXXXXXX}

\acmISBN{978-1-4503-XXXX-X/2018/06}

\begin{document}

\lstdefinelanguage{Solidity}{
  keywords={modifier, require, contract, function, using, for, public, library, internal, pure, returns,external, payable, ifAdmin},
  ndkeywords={msg, sender, owner},
  ndkeywordstyle=\color{black},
  comment=[l]{//},
  commentstyle=\color{gray}\itshape,
  stringstyle=\color{red},
  basicstyle=\ttfamily,
  sensitive=true,
  morestring=[b]"
}

\lstset{
  xleftmargin=.01\textwidth, xrightmargin=.015\textwidth,
  language=Solidity,
  frame=single,
  numbers=left,
  numberstyle=\tiny\color{gray},
  numbersep=4pt,
  stepnumber=1,
  tabsize=2,
  showstringspaces=false,
  breaklines=true,
  breakatwhitespace=true,
  basicstyle=\ttfamily\footnotesize,
  keywordstyle=\color{blue}\bfseries,
  commentstyle=\color{gray}\itshape,
  stringstyle=\color{red},
}

\title{Automated Vulnerability Injection in Solidity Smart Contracts:\\ A Mutation-Based Approach for Benchmark Development}


\author{Gerardo Iuliano, Luigi Allocca, Matteo Cicalese, Dario Di Nucci}
\affiliation{%
  \institution{University of Salerno}
  \city{Fisciano (SA)}
  \country{Italy}}
\email{geiuliano@unisa.it, l.allocca8@studenti.unisa.it, mcicalese@unisa.it, ddinucci@unisa.it}





\renewcommand{\shortauthors}{Iuliano et al.}

\begin{abstract} 
The security of smart contracts is critical in blockchain systems, where even minor vulnerabilities can lead to substantial financial losses.
Researchers proposed several vulnerability detection tools evaluated using existing benchmarks. However, most benchmarks are outdated and focus on a narrow set of vulnerabilities. 
This work evaluates whether mutation seeding can effectively inject vulnerabilities into Solidity-based smart contracts and whether state-of-the-art static analysis tools can detect the injected flaws. We aim to automatically inject vulnerabilities into smart contracts to generate large and wide benchmarks.
We propose \muse{}, a tool to generate vulnerable smart contracts by leveraging pattern-based mutation operators to inject six vulnerability types into real-world smart contracts. We analyzed these vulnerable smart contracts using Slither, a static analysis tool, to determine its capacity to identify them and assess their validity. 
The results show that each vulnerability has a different injection rate. Not all smart contracts can exhibit some vulnerabilities because they lack the prerequisites for injection. 
Furthermore, static analysis tools fail to detect all vulnerabilities injected using pattern-based mutations, underscoring the need for enhancements in static analyzers and demonstrating that benchmarks generated by mutation seeding tools can improve the evaluation of detection tools.
\end{abstract}




\maketitle

\section{Introduction}
\label{sec:intro}

Smart contracts are self-executing programs operating on blockchain platforms like Ethereum~\cite{buterin_ethereum}. They enable decentralized applications to execute pre-defined terms and conditions autonomously, promoting trust between untrusted parties. With the advancement of blockchain technology, smart contracts have become indispensable in various domains, including digital payments and decentralized finance~\cite{de-fi}. However, their immutable nature, while foundational to the blockchain trust model, presents significant challenges when vulnerabilities are discovered~\cite{smarter_sc}. Unlike traditional software, deployed smart contracts cannot be modified or patched, making even minor flaws potentially catastrophic~\cite{attacks}.

The security of smart contracts is critical. Vulnerabilities such as \textit{Reentrancy}, \textit{Timestamp dependence}, \textit{Transaction Order Dependence} (TOD), \textit{Authorization through tx.origin}, and \textit{Unchecked external calls} have already led to significant financial losses, exemplified by the infamous DAO hack~\cite{dao_attack}. Researchers and developers have created a myriad of vulnerability detection tools. Often leveraging static or dynamic analysis, these tools aim to identify and mitigate potential defects before deployment and, in some cases, on-chain. Despite their advancements, these tools are not infallible. False positives~\cite{fp2}, false negatives, and the inability to detect complex vulnerability patterns limit their effectiveness, underscoring the need for robust evaluation.

One critical barrier to improving detection tools is the lack of comprehensive and diverse benchmarks~\cite{benchmark}. Existing datasets are often outdated, limited in size, or narrowly focused on specific vulnerabilities, leaving many tools untested against realistic scenarios. The most commonly used benchmarks~\cite{smartbugs_wild} comprise smart contracts written using the old Solidity version and affected by a subset of the known vulnerability. Other datasets often contain simplistic toy contracts~\cite{TSE}, like the SWC Registry. In addition, about 96\% of smart contracts present in datasets are involved in no more than five transactions~\cite{understendingSC}. Ren \textit{et al.}~\cite{ren} highlighted that tools should be evaluated using a comprehensive benchmark suite that integrates multiple vulnerability types. The absence of high-quality and updated benchmarks hinders progress in the field by affecting the ability to validate and improve detection tools.

To address this gap, we propose \muse{}, a mutation-based tool based on \textsc{SuMo}~\cite{sumo} to generate benchmarks by injecting vulnerabilities into smart contracts. By leveraging mutation operators designed around known vulnerability patterns, our approach systematically introduces faults into realistic scenarios, enabling the generation of contracts with vulnerabilities placed in both typical and unconventional yet valid locations, challenging detection tools to expand their scope and improve accuracy. Such versatility allows for evaluating detection tools against various scenarios, including edge cases.
The strength of our approach lies in its foundation on pattern-based mutation operators. These operators are designed to inject vulnerabilities wherever their corresponding patterns are identified, ensuring consistency and extensibility; as new vulnerabilities are identified, corresponding operators can be added with minimal effort. 
We manually validated mutation operators and evaluated the generated benchmarks using \textsc{Slither}~\cite{slither}, a state-of-the-art static analysis tool for smart contracts. 

The results reveal that each vulnerability has a different injection rate, and successful injection depends on whether the smart contract satisfies the necessary preconditions to adhere to vulnerability patterns. Detection outcomes reveal significant limitations of \textsc{Slither} in identifying vulnerabilities, especially when they are injected into unconventional or unexpected locations. These findings underscore the weaknesses of current static analysis tools and highlight the pressing need for more advanced detection techniques. Moreover, \muse{} allows to increase the benchmark size of above 840\%.
In conclusion, our study identifies the gaps in the static analyzer, offering a tangible solution for researchers and practitioners to enhance or generate new benchmarks to evaluate their detection tools. 
Our work provides the following contributions:

\begin{itemize}
    \item \muse{}, a mutation seeding tool to generate benchmarks injecting vulnerabilities in Solidity smart contracts using pattern-based mutation operators;
    \item an enhanced version of the dataset smartbugs-wild containing 350,493 vulnerable smart contracts;
    \item a list of weaknesses that affect the capabilities of \textsc{Slither} in detecting vulnerabilities.
\end{itemize}

\paragraph{Paper Structure} This paper is organized as follows. \Cref{sec:background} establishes the background and reviews related work. \Cref{sec:method} details the research method, including research questions and the mutation-based approach. \Cref{sec:results} presents the experimental results, while \Cref{sec:discussion} discusses key findings, implications, and their relevance. \Cref{sec:ttv} addresses threats to validity, and \Cref{sec:conclusion} concludes the paper and provides future research directions.

\section{Background and Related Work}
\label{sec:background}
This section introduces smart contracts, their vulnerabilities, and mutation testing. Furthermore, we provide some related work that is relevant to our study.

\subsection{Smart Contracts}
Smart contracts are self-executing programs designed to automatically enforce terms and conditions between untrusted parties~\cite{9667515}. Initially envisioned as a way to automate legal contracts, the rise of blockchain technology has transformed smart contracts into scripts that execute synchronously across nodes in a distributed ledger~\cite{8847638}.
On the Ethereum blockchain, smart contracts run within the Ethereum Virtual Machine (EVM), a Turing-complete, stack-based virtual machine that ensures isolated contract code execution. A smart contract is identified by a unique address, private storage, and a balance in Ether, and it contains executable code. When a transaction is sent to a contract address, it triggers the contract functionality, providing invocation data and paying transaction fees using Gas~\cite{9667515}.

\subsection{Smart Contract Vulnerabilities}
Ethereum smart contracts are prone to various vulnerabilities unique to blockchain technology. Reentrancy is one of the most critical~\cite{kushwaha, reguard}, as evidenced by the infamous DAO hack, where an attacker repeatedly called back into a contract to drain its funds. Another major issue is the misuse of transaction origin for authorization, which attackers can easily spoof to gain unauthorized access.
Timestamp manipulation by miners is another concern, allowing them to alter timestamps and compromise the security of critical contract functions~\cite{mense, contractfuzzer}. Transaction-ordering dependence (TOD), where the order of transactions is unpredictable, can be exploited to unfairly manipulate outcomes, such as reducing rewards before submitting a valid solution~\cite{8726833, sayeed, etherfuzz}.
External calls also pose significant risks. Attackers can exploit these calls to execute malicious code. Furthermore, failing to handle a function return value properly may enable attackers to drain contract balances. Denial of Service (DoS) attacks can arise in several ways~\cite{smartscan}, such as through costly external calls or inefficient looping behavior.
Smart contract development differs fundamentally from traditional software programming. Vulnerabilities in smart contracts deployed on public blockchains are particularly challenging to fix due to the immutable nature of blockchain systems. While some traditional security techniques are applicable, smart contracts introduce unique challenges, and many vulnerabilities arise from the distinctive characteristics of blockchain technology. 

\subsection{Mutation Testing}
Mutation testing is a software testing technique to evaluate the effectiveness of test cases by intentionally introducing small changes, called mutants, into the source code to simulate potential faults or errors. The primary goal is to assess whether the existing test cases can detect these changes, thereby measuring the fault-detection capability of a test suite~\cite{PAPADAKIS, mutationTesting}. This process ensures software is rigorously validated, improving its reliability and reducing the likelihood of undetected faults in production.

Mutation testing in Solidity applies the same principles as traditional mutation testing. Still, it focuses on the unique characteristics of smart contracts, injecting Solidity-specific faults to evaluate the effectiveness of vulnerability detection and test suite robustness. In the literature, some tools have been proposed for this purpose.
Chapman \textit{et al.} proposed \textsc{Deviant}~\cite{Deviant}, a mutation testing tool designed for Solidity smart contracts. It automatically generates mutated versions of a given Solidity project and runs them against existing test suites to assess their effectiveness. \textsc{Deviant} includes mutation operators that cover Solidity-specific features based on a Solidity fault taxonomy and traditional programming constructs to simulate faults. The authors used \textsc{Deviant} to evaluate the test effectiveness of three Solidity projects. Their findings show that achieving high statement and branch coverage in Solidity does not guarantee strong code quality. This study provides valuable insights for Solidity developers, emphasizing the need for more rigorous testing to minimize financial risks.
Ivanova and Khritankov presented \textsc{RegularMutator}~\cite{RegularMutator}, a tool for improving the reliability of smart contracts written using Solidity language. \textsc{RegularMutator} implements language-specific operators that correspond to common errors made during the development of smart contracts. Injection of mutations in program code is implemented using regular expressions. The tool demonstrated its effectiveness in testing large-scale smart contract projects. The study concluded that mutation analysis provides a more reliable measure of test suite quality than traditional test line coverage.
Barboni \textit{et al.} proposed \textsc{SuMo}~\cite{sumo_journal}, a mutation testing tool designed for Solidity smart contracts, incorporating 25 Solidity-specific mutation operators alongside 19 traditional ones. It enables mutation testing on Solidity projects to assess test effectiveness. \textsc{SuMo} was later extended by \textsc{ReSuMo}~\cite{resumo}, which introduces a regression mutation testing approach. \textsc{ReSuMo} employs a static, file-level technique to selectively mutate a subset of smart contracts and rerun only relevant test cases during regression testing. After each mutation testing run, \textsc{ReSuMo} incrementally updates its results by leveraging test outcomes from previous program revisions, improving efficiency and reducing redundant computations.

\subsection{Related Work}
Bug injection is a testing method widely studied in traditional software programs; however, only a few studies have addressed its application to smart contracts using mutation security testing.

Ghaleb and Pattabiraman proposed \textit{SolidiFI}~\cite{solidifi}, an automated and systematic approach to evaluate static analysis tools. The tool injects bugs into a smart contract to introduce the targeted vulnerabilities and then checks the generated buggy contracts using the static analysis tools. The tool injects bugs by adding vulnerable code snippets, code transformations, and weakening security mechanisms. We extended the code transformation approach, focusing on more vulnerabilities and a larger dataset to experiment. We also validated the mutation seeding tool to provide a tool to generate new vulnerable smart contracts starting from an initial set.

Chu \textit{et al.}~\cite{SGDL_rw} introduced \textit{SGDL} (Smart Contract Vulnerability Generation with Deep Learning), an approach to create authentic and diverse vulnerability datasets for smart contracts. \textit{SGDL} combines generative adversarial networks (GANs) with static analysis to extract syntactic and semantic information from contracts. Using this information, it generates realistic vulnerability fragments and injects them into smart contracts via an abstract syntax tree, ensuring syntactic correctness.
The approach depends on a labeled dataset and researchers' expertise in vulnerabilities to train the GANs. However, comprehensive datasets for various vulnerabilities remain scarce or are not openly accessible for academic research. Consequently, \textit{SGDL} focus is limited to specific vulnerabilities.
In contrast, our mutation-based approaches rely on predefined patterns and manually designed mutation operators. New operators do not depend on datasets but on vulnerability patterns, making the approach easy to extend to other vulnerabilities.

Hajdu \textit{et al.}~\cite{FaultInjection} conducted a study using software-implemented fault injection (\textit{SWIFI}) to evaluate the dependability of permissioned blockchain systems in the presence of faulty smart contracts. They introduced general software and blockchain-specific faults into smart contract code to assess their impact on system reliability and integrity. They also investigated the effectiveness of formal verification and runtime protection mechanisms in detecting and mitigating these faults.
The authors used Hyperledger Fabric and evaluated 15 smart contracts, each tested in three versions: a base version, a version with extensive protections, and a version without protections. Faults were injected into the unprotected versions, resulting in 651 faulty variants. The findings revealed that formal verification and runtime protections complement built-in platform checks but cannot detect all faults.
Our study focused on a significantly larger dataset of faulty smart contracts and targeted critical smart contract vulnerabilities. 
Additionally, the goal was distinct, focusing on generating benchmarks that can be used to validate and improve the effectiveness of vulnerability detection tools.

Regarding benchmark generation, other techniques have been explored in the literature, such as analyzing audit reports and leveraging large language models (LLMs).
On the one hand, Zheng \textit{et al.}~\cite{TSE} created a large-scale dataset of SWC weaknesses from real-world DApp projects. They recruited 22 participants to analyze 1,199 open-source audit reports from 29 security teams, identifying 9,154 weaknesses. Their work resulted in two distinct datasets.
The DAppSCAN-Source dataset contains 39,904 Solidity files with 1,618 SWC weaknesses. However, these files may not be directly compilable for automated analysis. To address this, the authors developed a tool that automatically identifies dependency relationships within DApp projects and resolves missing public libraries.
The DAppSCAN-Bytecode dataset includes 6,665 compiled smart contracts containing 888 SWC weaknesses. Evaluation results showed that existing detection tools perform poorly on these datasets, highlighting the need for future research to focus on real-world smart contract datasets rather than simplistic toy contracts.
On the other hand, Daspe \textit{et al.}~\cite{daspe} utilized large language models (LLMs) to generate a dataset of Solidity smart contracts. To guide the LLM during the generation process, they adopted an approach inspired by Test-Driven Development (TDD)~\cite{tdd}. Each prompt was submitted to an LLM, producing Solidity code that was then parsed. The Solidity compiler verified the syntax, and if the contract compiled successfully, Slither was applied for static analysis to detect vulnerabilities. Next, they created a project containing the generated contract and functional tests, which were then executed. They collected the compilation status, vulnerability report, and functional test results for each prompt. After multiple generations, they conducted an evaluation based on prompt complexity and the model used. The study found that LLMs struggled with increased complexity, demonstrating low accuracy as contract intricacy grew. Most compilation errors arose from incorrect type usage, like strings and arrays, and issues related to the payable/call pattern.

The work described above highlights the challenges faced and the different approaches used in literature to generate benchmarks. Our work bridges some of the gaps by offering a possible solution.
\begin{figure*}[ht]
    \centering
    \caption{Summary of the Research Method.}
    \Description{Research method}
    \label{fig:research_method}
    \includegraphics[width=0.75\linewidth]{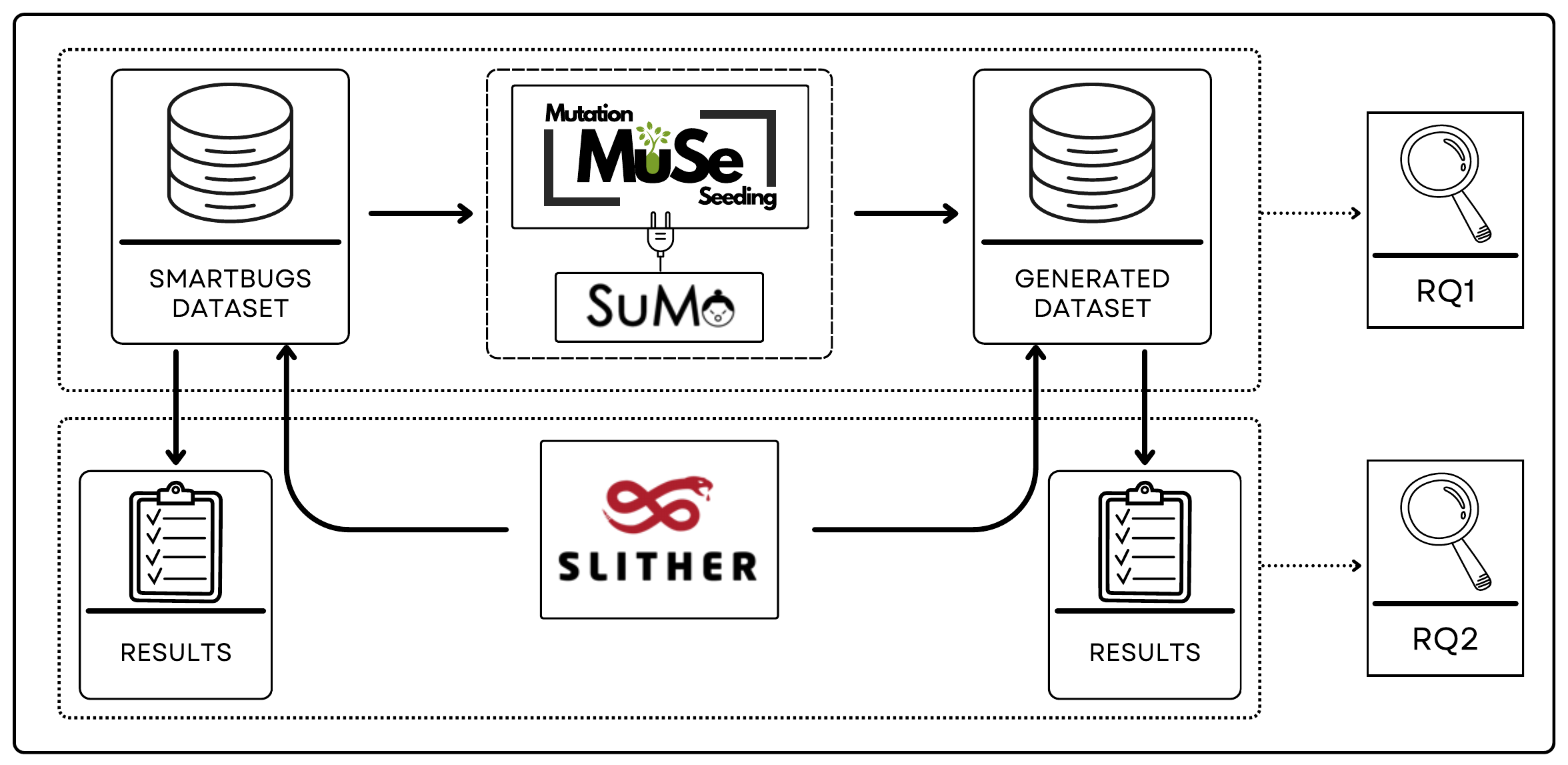}
\end{figure*}
\section{Research Method}
\label{sec:method}

The following section presents the details of the study, highlighting the main goal and its related research questions. 

\goal{Our goal is to automatically inject vulnerability in smart contracts to generate large and wide benchmarks that researchers and developers can use to improve detection tool evaluation.}

To achieve this goal, we developed \muse{}, a tool to generate vulnerable smart contracts that are challenging to detect automatically. Inspired by mutation testing~\cite{PAPADAKIS, mutationTesting}, we extended the \textsc{SuMo} mutation testing tool~\cite{sumo} to mutate Solidity smart contracts into vulnerable mutants.
To this end, we implemented mutation operators designed to inject known vulnerabilities by modifying the smart contracts appropriately. Afterward, we analyzed to what extent the injected vulnerabilities are challenging to detect by observing how well static analyzers can detect them.
We leveraged \textsc{Slither}~\cite{slither}, a popular static analyzer to detect vulnerabilities in Solidity, and compared the results achieved by running it before and after the mutation phase, assessing the presence of the injected vulnerabilities and the performance of \textsc{Slither} in detecting them.

Our motivation is the critical role that smart contract security plays in the blockchain and the limitations of existing vulnerability detection tools in handling complex vulnerabilities. By combining mutation security testing with static analysis, this work seeks to provide empirical evidence of the strengths and weaknesses of tools like \textsc{Slither}, offering insights that can drive the development of more robust security solutions and methodologies for smart contracts.
\Cref{fig:research_method} depicts our research method.

\begin{table*}[ht]
    \centering
    \footnotesize
    \caption{Vulnerabilities injected into smart contracts by \muse{}.}
    \begin{tabular}{l p{8cm} c c}
        \hline \textbf{Vulnerability} & \textbf{Description} & \textbf{Slither's Detector} & \textbf{Operator}\\ \hline
        Unchecked low-level call return value & Low-level calls return \textit{false} on failure instead of throwing exceptions, risking critical vulnerabilities if unchecked. & \href{https://github.com/crytic/slither/wiki/Detector-Documentation#unchecked-low-level-calls}{\texttt{unchecked-lowlevel}} & UC\\
        Unchecked send & The send function returns \textit{false} on failure without throwing an exception, risking vulnerabilities if unchecked. & \href{https://github.com/crytic/slither/wiki/Detector-Documentation#unchecked-send}{\texttt{unchecked-send}} & US\\
        Authentication via tx.origin & Using \textit{tx.origin} for authorization risks vulnerabilities if an authorized account interacts with a malicious contract. & \href{https://github.com/crytic/slither/wiki/Detector-Documentation#dangerous-usage-of-txorigin}{\texttt{tx-origin}} & TX\\
        Unused return & The return value of an external call is not stored in a local or state variable. & \href{https://github.com/crytic/slither/wiki/Detector-Documentation#unused-return}{\texttt{unused-return}} & UR\\
        Multiple calls in a loop & Calls inside a loop might lead to a denial-of-service attack. & \href{https://github.com/crytic/slither/wiki/Detector-Documentation#calls-inside-a-loop}{\texttt{calls-loop}} & CL\\ 
        Delegatecall to untrusted callee & \textit{Delegatecall} executes the code at the target address in the context of the calling contract. It allows a SC to load code dynamically from a different address. & \href{https://github.com/crytic/slither/wiki/Detector-Documentation#controlled-delegatecall}{\texttt{controlled-delegatecall}} & DTU\\ \hline
    \end{tabular}
    \label{tab:vuls}
\end{table*}

As shown in \Cref{tab:vuls}, we selected six vulnerabilities~\cite{iulianoSLR, ramederSLR, openscv} to inject based on their relevance in the literature and the ability of \textsc{Slither} to detect them with at least medium confidence: \textit{Unchecked call return value}, \textit{Unchecked send}, \textit{Authentication through tx.origin}, \textit{Delegatecall to untrusted callee}, and \textit{Unused return}. These vulnerabilities are among the top 15 most discussed in the literature, as highlighted in the work of Zaazaa Oualid and El Bakkali Hanan~\cite{Zaazaa}.
Based on our goal, we formulate these research questions (RQs):

\examplebox{\faSearch \hspace{0.2mm} \textbf{RQ\textsubscript{1}.} To what extent are the mutation operators implemented in \muse{} generalizable?}

\textbf{$RQ_1$} allows us to assess the feasibility of introducing vulnerabilities into real-world smart contracts and understand the generalizability of the mutation operators to inject a vulnerability.


\examplebox{\faSearch \hspace{0.2mm} \textbf{RQ\textsubscript{2}.} How can the mutants injected by \muse{} be detected through static analysis?}

\textbf{$RQ_2$} analyzes the performance of static analyzer when detecting vulnerabilities injected into contracts. The goal of applying various patterns to introduce vulnerabilities is twofold. On the one hand, we aim to identify whether the injected vulnerabilities are challenging to detect. On the other hand, we want to evaluate whether a static analysis tool can identify these specific patterns, highlighting its strengths and weaknesses.

\subsection{Data Collection}
To answer our research questions, we used the real-world dataset smartbugs-wild\footnote{https://github.com/smartbugs/smartbugs-wild}, which contains 47,398 smart contracts extracted from the Ethereum network that have at least one transaction.
We ran \textsc{Slither} using SmartBugs~\cite{smartbugs_tool}, which allows us to parallelize its execution using several Docker images and easily parse the results.

\subsection{Mutation Operators}
We extended SuMo~\cite{sumo}, a mutation testing tool, by creating new mutation operators focusing on security. The tool uses the \textit{solidity-parser-antlr}\footnote{https://github.com/solidity-parser/parser}, a parser built from a robust ANTLR4 grammar, which generates an Abstract Syntax Tree of the code based on the Solidity grammar. We implement a mutation operator for each vulnerability to inject and leverage the parser to identify injection patterns where the vulnerabilities can be injected. \Cref{tab:vuls} maps the implemented mutation operators, the vulnerability they inject, and the detector \textsc{Slither} uses to identify them. 

\paragraph{UC Operator} To inject the \textit{Unchecked low-level call return value}, we identify the possible instructions or statements to mutate. We identified all the low-level call functions. We mutated the contract by removing the controls whenever the return value of the functions was checked using the require function or an if statement.

\begin{tcolorbox}[colback=gray!5!white, colframe=white, left=6pt, right=0pt, top=0pt, bottom=0pt]
\begin{lstlisting}[frame=none]
// Before UC mutation
function withdraw(uint amount) public {
  require(msg.sender.call.value(amount)());
}
// After UC mutation
function withdraw(uint amount) public {
  msg.sender.call.value(amount)();
}
\end{lstlisting}
\end{tcolorbox}

\paragraph{US Operator} To inject \textit{Unchecked send}, we identified all the send functions and removed any control on the return value.

\begin{tcolorbox}[colback=gray!5!white, colframe=white, left=6pt, right=0pt, top=0pt, bottom=0pt]
\begin{lstlisting}[frame=none]
// Before US mutation
function sendEth(address payable giftee) public {
  if (!giftee.send(1 ether)) {
    revert("Send failed");
  }
}
// After US mutation
function sendEth(address payable giftee) public {
  giftee.send(1 ether) 
}
\end{lstlisting}
\end{tcolorbox}

\paragraph{TX Operator} \textit{Authentication through tx.origin} was injected substituting the \texttt{msg.sender} variable with \texttt{tx.origin}. We mutate the contracts when the \texttt{msg.sender} variable is used in a binary operation like \texttt{``==''} to check the ownership of the contract or a specific address having some privileges or access to the asset.

\begin{tcolorbox}[colback=gray!5!white, colframe=white, left=6pt, right=0pt, top=0pt, bottom=0pt]
\begin{lstlisting}[frame=none]
// Before TX mutation
modifier onlyOwner() {
  require(msg.sender == owner, "No owner"); _;
}
// After TX mutation
modifier onlyOwner() {
  require(tx.origin == owner, "No owner"); _;
}
\end{lstlisting}
\end{tcolorbox}

\paragraph{UR Operator} \textit{Unused return} was injected in two cases. First, when a binary operation like \texttt{``=''} assigns the return value of a function to a variable.
Second, when a variable is declared and then initialized with the return value of a function.
In both cases, we removed the left side of the assignment and left the call function.

\begin{tcolorbox}[colback=gray!5!white, colframe=white, left=6pt, right=0pt, top=0pt, bottom=0pt]
\begin{lstlisting}[frame=none]
// Before UR mutation
function addNumbers(uint256 a, uint256 b) public {
  c = SafeMath.add(a, b); 
}
// After UR mutation
function addNumbers(uint256 a, uint256 b) public {
  SafeMath.add(a, b); 
}
\end{lstlisting}
\end{tcolorbox}

\paragraph{CL Operator} Multiple calls in a loop were injected by wrapping call, send, and transfer functions in a statement of 1,000 loops implemented using the \texttt{``for''} construct.

\begin{tcolorbox}[colback=gray!5!white, colframe=white, left=6pt, right=0pt, top=0pt, bottom=0pt]
\begin{lstlisting}[frame=none]
// Before CL mutation
function payMember(address payable member) public {
  require(member.send(0.1 ether);
}
// After CL mutation
function payMember(address payable member) public {
  for (uint256 i = 1; i <= 5; i++) {
    require(member.send(0.1 ether);
  }
}
\end{lstlisting}
\end{tcolorbox}

\paragraph{DTU Operator} \textit{Delegatecall to untrusted callee} was injected by introducing a new address variable and a new function that allows users to replace the address variable with another. Each use of \texttt{delegatecall} was mutated, replacing the address used to delegate with the new personalizable address, which can be set to malicious.


\begin{tcolorbox}[colback=gray!5!white, colframe=white, left=6pt, right=0pt, top=0pt, bottom=0pt]
\begin{lstlisting}[frame=none]
// Before DTU mutation
function setFalseValue(address _address) public {
  require(_address.delegatecall(
    abi.encodeWithSignature("setFalse(uint256)")));
}
// After DTU mutation
address public delegate;
function setDelegate(address _delegate) public { 
  delegate = _delegate; 
}
function setFalseValue(address _address) public {
  require(delegate.delegatecall(
    abi.encodeWithSignature("setFalse(uint256)")));
}
\end{lstlisting}
\end{tcolorbox}

\subsection{\muse{} Validation}
To ensure the validity of our results, we manually validated \muse{} on a statistically significant subset of mutated smart contracts with a 95\% confidence level and a 5\% margin error. The tool generated 350,716 mutated contracts, from which we randomly selected a subset of 384 smart contracts, representing a statistically significant sample size.
Our validation process involved manually analyzing each contract to verify whether SuMo injected the intended vulnerabilities correctly. The procedure consisted of these steps:

\begin{enumerate}
\item \textit{Compilation for Syntactical Correctness.}
Each selected smart contract was compiled to ensure its syntactical correctness.
\item \textit{Comparison with SuMo Logs.}
We compared the logs provided by SuMo, which detail the applied mutation type and the lines of code affected, against the corresponding mutated smart contracts.
\item \textit{Pattern Adherence Verification.}
We verified that the points where vulnerabilities were injected adhered to the patterns defined by the mutational operator. This step ensured that SuMo identified the correct lines of code and statements for injecting vulnerabilities.
\item \textit{Modification Assessment.}
We examined SuMo's modifications or additions to the original code to determine whether the injected vulnerabilities were accurately implemented.
\end{enumerate}

A mutation was marked as correctly injected if (i) the mutated contract contained the new lines of code introducing the target vulnerability or (ii) existing lines of code were altered to render the smart contract vulnerable to the intended issue.

The validation results show that our tool failed to inject vulnerabilities in 20 out of 384 smart contracts, or 5.21\% of cases.  
The main causes of these failures are exceptional cases that the mutational operator does not adequately handle. One common issue arises when the mutated statement contains a semicolon (\texttt{``;''}). Strings including a semicolon within the mutated statement can interfere with the operator logic, leading to unintended code truncation. As a result, the generated mutant may have incorrect syntax, rendering the contract uncompilable.
Another cause of injection failure is conflicts between the scopes of contract variables and the variables introduced by the mutation. For example, the CL operator injects a for loop that uses the variable \texttt{uint i} for iteration. If the mutation is applied to a statement declaring a \texttt{uint i} variable, the compiler cannot differentiate between the two variables, leading to a compilation error.
In all other cases, the mutation process successfully injects the vulnerability without issues, producing valid mutants.

\subsection{Replication Package}
We have made \muse{} publicly available on GitHub\footnote{\url{https://anonymous.4open.science/r/MuSe/}}, allowing researchers and practitioners to replicate our study or utilize the tool for their purposes. Additionally, we have uploaded the sample used to manually validate the mutation operators.\footnote{\url{https://figshare.com/s/5473d31f2aead13d2fa8}}
\section{Empirical Results}
\label{sec:results}

We ran \muse{} on the SmartBugs-wild dataset, applying the six previously described mutation operators to each contract. Starting from 47,398 smart contracts, we generated 350,493 vulnerable ones. \muse{}  mutates a contract each time it matches the pattern of a mutation operator. A contract could exhibit more than a pattern. \Cref{tab:results_rq1} shows the number of mutants generated by each operator and highlights the number of contracts suitable to be mutated by each operator.
\muse{} mutated 41,337 out of 47,398 smart contracts, about 87\% of cases. The remaining 6,061 smart contracts were not mutated for two reasons: (i) the absence of any patterns used by mutation operators in 5,990 smart contracts and (ii) the invalid content of the files for 71 of them, e.g., a JSON representation of the smart contract instead of well-formatted Solidity code.


\subsection{RQ1. To what extent are the mutation operators implemented in \muse{} generalizable?}
To answer RQ\textsubscript{1}, we observed the number of smart contracts that could be correctly mutated. We aimed to understand how many pattern occurrences applied by each mutation operator could be injected in real-world scenarios.

\begin{table}[h]
    \centering
    \caption{Results for mutation operators ordered by injection rate, total number of generated mutants, and average injection rate of \muse{}.}
    \label{tab:results_rq1}
    \begin{tabular}{lrrr}
    \hline
        Operator & \# Mutated SCs & \# Mutants & Injection Rate\\ \hline
        UR  & 33,910 & 213,912 & 71.50\%\\
        TX  & 32,250 & 65,825 & 68.00\%\\
        CL  & 26,604 & 61,687 & 56.00\%\\
        UC  & 4,094 & 4,992 & 8.60\%\\
        US  & 2,248 & 3,928 & 4.70\%\\
        DTU & 113 & 149 & 0.23\%\\ \hline
        -  &  -  & 350,493 & 34.83\%\\
        \hline
    \end{tabular}
\end{table}

As shown in \Cref{tab:results_rq1}, the most injectable vulnerability is the \textit{unused return}, with an injection rate of 71.5\%.
The patterns used to inject vulnerabilities are not only common but also frequently encountered in smart contracts, highlighting their prevalence in typical contract design.
On average, each smart contract contains six occurrences of these patterns, reflecting their foundational role in contract development. Patterns such as assignments, declarations, and initializations are essential building blocks in smart contract programming. However, their prevalence also increases the likelihood of vulnerabilities arising from improper or unintended usage. 

The second vulnerability, \textit{authorization via tx.origin}, is injectable in a real-world scenario in 68\% of cases. The high injection rate respects the frequency of the pattern of the TX operator in the smart contracts. As described in \Cref{sec:method}, the TX operator mutates a contract when the variable \texttt{msg.sender} is used to check the ownership of a contract or the privilege of an address on the asset. 

The injection rate for \textit{multiple calls in a loop} is 56\%, showing only just over half of the contracts involve the use of a \texttt{call}, \texttt{send}, or \texttt{transfer} function. The moderate injection rate highlights that these functions are commonly employed in contracts but not excessively frequent. In addition, 8\% of the unmutated smart contracts already contained the vulnerability.

\textit{Unchecked low-level call return value} and \textit{unchecked send} follow with injection rates of 8.6\% and 4.7\%, respectively. 
The frequency of a \texttt{call} function is almost double that of a \textit{send} function. It is generally better to use \textit{call} function than \texttt{send} but with some important security and implementation considerations. The \texttt{call} function is more flexible and allows specifying the amount of gas sent and calls with data. Nevertheless, it is more vulnerable to reentrancy attacks because it enables the receiving contract to execute arbitrary code.

Finally, \textit{delegatecall to untrusted callee} has the lowest injection rate. The use of the \texttt{delegatecall} function is relatively rare but not negligible. It is mainly limited to specific use cases that require advanced behavior, like implementing the Proxy pattern.

By analyzing the injection rate, it is possible to see that some contracts do not have the necessary conditions to inject a given vulnerability. Contracts that lack specific patterns, constructs, or known Solidity functions are intrinsically safe from vulnerabilities that try to exploit these elements. Overall, mutation operators can increase the size of a dataset by 840\% by creating new vulnerable versions of smart contracts. 

\examplebox{\faKey \hspace{0.2mm} \textbf{RQ\textsubscript{1} Summary.} The results highlight that the patterns needed to inject ``unused return'' (71.5\%), ``authorization via tx.origin'' (68\%), and ``multiple calls in a loop'' (56\%) are common, whereas those related to ``unchecked low-level call return values'' (8.6\%) and ``unchecked send'' (4.7\%) are less prevalent. The pattern for ``delegatecall to untrusted callee'' is rare (0.23\%) because the delegatecall function is used infrequently and only in specific locations.}

\subsection{RQ2. How can the mutants injected by \muse{} be detected through static analysis?}

To answer RQ\textsubscript{2}, first, we performed an initial detection with \textsc{Slither} on the smartbugs-wild dataset and collected the findings for each contract to have a baseline. Then, we ran \textsc{Slither} on the mutants generated by \muse{}. 
Under the assumption that the mutation operator correctly injects the vulnerability, we compared the detection results before and after the mutation. \textsc{Slither} correctly detects a mutant if it is labeled as vulnerable to the type of injected vulnerability. In the case where the contract is already vulnerable and has been mutated, we checked whether, in addition to the pre-existing vulnerability, the injected vulnerability had also been detected by analyzing the lines of code related to the injected vulnerability.

\textsc{Slither} successfully analyzed 335,234 mutants out of the 350,493 generated. Similarly, the execution on the smartbugs-wild dataset failed on 2,700 smart contracts out of 47,398. 
After excluding the failed executions, we mapped the results achieved by \textsc{Slither} on the original contracts with the mutated ones, if present, and compared the results for further analysis.

\begin{table}[ht]
    \centering
    \caption{Detection rate of injected vulnerability and overall performance of Slither on the six considered vulnerabilities.}
    \label{tab:results_rq2}
    \begin{tabular}{lrrrr}
    \hline
        Vulnerability & TP & FN & Recall & FNR\\ \hline
        UC & 4,876 & 0 & 1.000 & 0.000\\
        US & 3,570 & 0 & 1.000 & 0.000\\
        CL & 45,261 & 10,563 & 0.810 & 0.189\\
        UR & 124,858 & 81,184 & 0.605 & 0.394\\
        TX & 21,765 & 42,937 & 0.336 & 0.663\\
        DTU & 15 & 134 & 0.100 & 0.899\\ \hline
        -   & 200,345 & 134,818 & 0.597 & 0.402\\
        \hline
    \end{tabular}
\end{table}

As shown in \Cref{tab:results_rq2}, the results achieved by \textsc{Slither} against the \textit{unchecked low-level call return value} (UC) and \textit{unchecked send} (US) vulnerabilities are surprisingly high. The tool detected all mutants with the injected vulnerability, showing a recall value equal to 1.00 in both cases. The two vulnerabilities are conceptually similar, which shows that \textsc{Slither} implements strong detectors to check whether the return values of the call and send functions are handled correctly. \textsc{Slither} also performs very well in detecting \textit{multiple calls in a loop} (CL), with a recall of 0.81. While the detection is strong, a noticeable portion of vulnerabilities remains undetected, indicating room for improvement in the detection mechanism of this vulnerability. Performance slows down on \textit{unused return} (UR) with a recall value of 0.63, suggesting that the detection mechanism of these vulnerabilities might be less robust or prone to specific limitations. Performance deteriorates on \textit{authorization via tx.origin} (TX), 0.33 of recall, pointing out significant gaps in detection capabilities for this category. The worst result is detecting \textit{delegatecall to untrusted callee} (DTU) with a recall value of 0.10.

Overall, \textsc{Slither} achieved a recall value of 0.597. The recall value indicates that while \textsc{Slither} effectively identifies certain vulnerabilities, it fails to detect a significant portion (40.2\%) of the injected vulnerabilities. The result underscores the need to improve static analysis tools or complement them with additional detection techniques to enhance their accuracy and reduce false negatives.


\examplebox{\faKey \hspace{0.2mm} \textbf{RQ\textsubscript{2} Summary.} \textsc{Slither} performs very well in some cases but inconsistently across vulnerability types. While it excels in detecting simple vulnerabilities like \textit{unchecked low-level call return} value and \textit{unchecked send}, it struggles with more complex vulnerabilities like \textit{authorization via tx.origin} and \textit{delegatecall to untrusted callee}. Finally, \textsc{Slither} detected the 59.7\% of the injected vulnerabilities using \muse{}.}

\section{Discussions and Limitations}
\label{sec:discussion}

In this section, we analyze the false negatives resulting from running Slither on the mutated smart contracts. We also discuss the limitations of the study, offering insights about our mutation-based approach to injecting vulnerabilities.

\subsection{False Negative Analysis}
We analyzed false negatives to extract information about the errors achieved by Slither in detecting vulnerabilities. The way we conducted the experiment allows us to extract only true positives (TP) and false negatives (FN). The mutational operators, validated as described in Section 3, inject the target vulnerability. Having Slither results before and after the mutation, we can analyze whether the tool detects the injected vulnerability (TP) or fails detection (FN).

\paragraph{Authorization via tx.origin (TX)} For this vulnerability, the FNR is 0.663. We found some patterns that Slither does not check when detecting this vulnerability type. The most relevant and alarming is related to the Solidity modifier. A Solidity modifier is a reusable function that encapsulates and enforces reusable logic, such as access control or precondition checks, simplifying code and improving maintainability. \Cref{fig:incorrectModifier} shows the incorrect implementation of a modifier used to check the contract owner using \texttt{tx.origin} and that Slither cannot label as vulnerable.


\begin{figure}[h]
\centering
\caption{Incorrect modifier to restrict owner's access.}
\label{fig:incorrectModifier}
\begin{tcolorbox}[colback=gray!5!white, colframe=white, left=6pt, right=0pt, top=0pt, bottom=0pt]
\begin{lstlisting}[
frame=none]
modifier onlyOwner() {
  require(tx.origin == owner); _;
}
\end{lstlisting}
\end{tcolorbox}
\end{figure}

Another scenario that Slither fails to detect is when the clause to check the contract ownership is composed of several conditions combined using AND/OR operators. For example, \Cref{fig:clause_and_or} shows the \texttt{addPartner} function, which allows the caller to add a new \texttt{partner}. The clause in the \texttt{require} function ensures that only authorized users (\texttt{\_dev} or \texttt{\_owner}) can call this function. The \textit{TX operator} mutated the second occurrence of \texttt{msg.sender}, replacing it with \texttt{tx.origin}. Using \texttt{tx.origin} for authorization introduces a security risk because it refers to the address that initiated the transaction, even if there were intermediate contract calls.


\begin{figure}[h]
\centering
\caption{Clause with two conditions combined using OR.}
\label{fig:clause_and_or}
\begin{tcolorbox}[colback=gray!5!white, colframe=white, left=6pt, right=0pt, top=0pt, bottom=0pt]
\begin{lstlisting}[frame = none]
function addPartner(address _partner) public {
  require((msg.sender == _dev) || (tx.origin == _owner));
  exchangePartners[_partner] = true;
}
\end{lstlisting}
\end{tcolorbox}
\end{figure}

\paragraph{Unused return (UR)} \textsc{Slither} achieved a false negative rate of 0.394 on this vulnerability. Analysis of false negatives produced an interesting finding. The unused return is correctly detected by \textsc{Slither} when the function being called in the contract is a library function; see \Cref{fig:detected}. However, the detection fails when the function called is inherited from a contract; see \Cref{fig:notDetected}.


\begin{figure}[h]
\centering
\caption{Two examples of unused return.}
\begin{subfigure}[b]{1\linewidth}
\caption{Unused return detected at line 10.}
\label{fig:detected}
\begin{tcolorbox}[colback=gray!5!white, colframe=white, left=6pt, right=0pt, top=0pt, bottom=0pt]
\begin{lstlisting}[frame=none]
library SafeMath {
  function add(uint256 a, uint256 b) internal pure returns (uint256) {
    uint256 c = a + b;
    require(c >= a, "addition overflow");
    return c;
  }
}
contract SafeMathExample {
  function addNumbers(uint256 a, uint256 b) public {
    SafeMath.add(a, b); 
  }
}
\end{lstlisting}
\end{tcolorbox}
\end{subfigure}
\hfill
\begin{subfigure}[b]{1\linewidth}
\caption{Unused return not detected at line 10.}
\label{fig:notDetected}
\begin{tcolorbox}[colback=gray!5!white, colframe=white, left=6pt, right=0pt, top=0pt, bottom=0pt]
\begin{lstlisting}[frame=none]
contract SafeMath {
  function add(uint256 a, uint256 b) internal pure returns (uint256) {
    uint256 c = a + b;
    require(c >= a, "addition overflow");
    return c;
  }
}
contract SafeMathExample is SafeMath{
  function addNumbers(uint256 a, uint256 b) public {
    SafeMath.add(a, b); 
  }
}
\end{lstlisting}
\end{tcolorbox}
\end{subfigure}
\end{figure}


\paragraph{Multiple calls in a loop (CL)} \textsc{Slither} achieved a false negative rate of 0.189 on this vulnerability. The mutated smart contracts are characterized by several aspects, like function visibility and the presence of modifiers. In addition, the mutation can be injected into the function bodies or in-depth, e.g., nested into an \texttt{if} statement. Finally, the statement that undergoes the mutation may be contained in another statement. Nevertheless, we could not find a recurrent pattern in the inconsistent behavior of the detection tool.

\paragraph{Delegatecall to untrusted callee (DTU)} The analysis of false negatives revealed two interesting aspects. On the one hand, we noticed that \textsc{Slither} fails to detect simple cases like the one shown in \Cref{fig:dtu_not_detected}. On the other hand, we noticed that most \texttt{delegatecall} functions are used in the constructor, which is invoked only at deployment time. Therefore, it is impossible to inject an instance of the vulnerability into the constructor that is exploitable.
In addition, the \texttt{delegatecall} is often rewritten using the assembly to create a custom function to delegate. 
Although the mutational operator showed a poor injection rate (0.23\%), the impact of this vulnerability is catastrophic if misused in the Proxy pattern. The Proxy pattern uses \texttt{delegatecall} to separate state and logic, allowing updates without losing contract data.


\begin{figure}[h]
\centering
\caption{Delegatecall to untrusted callee.}
\label{fig:dtu_not_detected}
\begin{tcolorbox}[colback=gray!5!white, colframe=white, left=6pt, right=0pt, top=0pt, bottom=0pt]
\begin{lstlisting}[frame=none]
address public delegate;
function setDelegate(address _delegate) public { 
  delegate = _delegate; 
}
function upgradeAndCall(address implementation, bytes calldata data) external payable ifAdmin {
  _upgradeTo(implementation);
  (bool success,) = delegate.delegatecall(data);
  require(success);
}
\end{lstlisting}
\end{tcolorbox}
\end{figure}

\subsection{Mutations and Side Effects}

This section analyzes the side effects that mutation testing could have on smart contracts.
We observed that injecting vulnerabilities through mutational operators often introduces side effects, like code smells. We relied on the \textsc{Slither} official documentation\footnote{https://github.com/crytic/slither/wiki/Detector-Documentation} to understand the functionality of its detectors and the labels they use for vulnerabilities. 

\paragraph{Unchecked low-level call return value (UC) and Unchecked send (US)} 
Both vulnerabilities are injected using the same approach. The differences are in the conditions used to inject them, but the type of mutation is quite similar. Indeed, the side effects that affect the injection of these vulnerabilities are the same and occur in 75\% of the cases. A \textit{deprecated-statement} occurs when outdated constructs are used in a contract, e.g., \texttt{throw} instead of \texttt{revert}. The two mutation operators, UC and US, remove checks on the return value of a function. When these checks are implemented using an \texttt{if} statement, the true branch typically reverts the execution of the function using a \texttt{throw} construct. By eliminating the check on the return value, the condition to trigger the \texttt{throw} is also removed, resulting in the removal of the \texttt{throw} from the mutated code. Since \texttt{throw} is deprecated, its removal eliminates using a deprecated statement in 3,772 cases. 

Analyzing cases (3,697) where the side effect is the \textit{missing-zero-check} reveals an interesting finding. This issue arises when no validation is performed to ensure that an address, either used as an argument or on which a function is invoked, is not the zero address (\textit{address(0)}). The zero address is often used as a burn address for tokens.
When the mutation operator modifies the condition by moving it outside an \texttt{if} statement, \textsc{Slither} identifies the presence of a missing zero check. Although the statement remains unchanged, its move outside the \texttt{if} condition allows \textsc{Slither} to detect an issue that should be detected regardless of its position in the code.

\paragraph{Authorization via tx.origin (TX)}
The mutation operator introduces some unexpected behavior in 70\% of the cases.
When a function performs critical mathematical operations on the contract balance, these operations must be signaled by throwing an event. Functions that perform mathematical operations on the contract balance and do not emit an event are labeled by \textsc{Slither} as \texttt{events-math}. When the TX operator mutates them, \textsc{Slither} stops labeling them vulnerable in 2,581 cases. The mutation does not involve the mathematical operations in the contract; therefore, they should continue to be labeled as \textit{events-math}. One possible explanation could be the absence of the \textsc{msg.sender} variable, which seems necessary for a math operation to be signaled by an event.

A similar case involves \texttt{events-access}, emitting an event whenever the \texttt{msg.sender} variable is used to change the owner of the contract. In 2,398 cases, it is understandable to eliminate this vulnerability when \texttt{msg.sender} is replaced with \texttt{tx.origin}.

The last side effect, called by \textsc{Slither} \textit{suicidal}, occurs 1,058 times and concerns using the \texttt{selfdestruct} function, which should be restricted to the contract owner. When the mutation replaces \texttt{msg.sender} with \texttt{tx.origin} in the modifier used to give access to \texttt{selfdestruct}, \textsc{Slither} correctly detects the misuse of \texttt{selfdestruct}. The detector applies the appropriate control, discussed in the False Negative Analysis section—specifically, verifying the contract owner using a modifier. However, the detection behavior appears inconsistent depending on the functionality implemented within the function. For instance, if the function includes the invocation of a well-known operation, such as \texttt{selfdestruct}, \textsc{Slither} verifies access to the function by analyzing the associated modifier. Conversely, \textsc{Slither} fails to check the modifier if the function does not include recognized features.
This inconsistency highlights a significant issue: regardless of the functionality implemented, an incorrect modifier implementation should always be detected, which is especially critical when the modifier is responsible for ensuring that the function invoker is the contract owner.

\paragraph{Unused return (UR)}
We noted that the mutation operator introduces other vulnerabilities or code smells alongside the injected vulnerabilities in 67\% of the cases. For example, in some cases, the operator splits a single statement containing declaration and initialization into two separate statements, one for declaring the variable and one for initializing it. The initialization is then made vulnerable if the return value of a function is used to initialize the variable. The return value is not assigned to the variable, thus creating an unused return and leaving the variable uninitialized. The side effect of this mutation is the creation or removal of some code smells. The mutation has introduced \textit{uninitialized-state} (27,890),  \textit{initialized-local} (97,932), or \textit{constable-states} (28,352). 
While not expected, these side effects are understandable, given the type of mutation introduced. Variables uninitialized due to the mutation introduce \textit{uninitialized-state} and \textit{uninitialized-local}, depending on the scope of the uninitialized variable. In cases where the variable affected by the mutation undergoes no other changes in the code, we have a \textit{constable-states}, a variable not declared constant. 

Interesting side effects are those that remove some vulnerabilities or code smells present before the mutation and that were removed by the mutation. For example, \textit{divide-before-multiply} (-3,913) is resolved when the return value of multiplication is not assigned to the variable in which the result of the previous division was saved. Another is \textit{incorrect-equality} (-3,851), which occurs when a variable to which the contract balance is assigned is used in a strict equality. Removing the contract balance assignment from the variable also removes the vulnerability. Then we have \textit{reentrancy-benign} (-5,340), \textit{reentrancy-no-eth} (-3,191), \textit{reentrancy-eth} (-830), and \textit{reentrancy-unlimited-gas} (-818), all of which are removed from the mutation because by not assigning the return value to a variable involved in reentrancy, it does not change state and does not create the preconditions for reentrancy. Additionally, \textit{controlled-array-length} was added and removed depending on the case. It was introduced in 371 mutants but removed in 790. It was eliminated when the array index was not initialized using the return value of a function and introduced when the index remained uninitialized.

Interestingly, some vulnerabilities or code smells emerged, although not directly connected to the mutation. For example, \textit{external-function} or \textit{timestamp} were introduced by the UR operator but appeared in different lines of code than those affected by the mutation. The mutated statement and the statement affected by the newly emerged vulnerability shared no common functions or variables.
We hypothesize that these instances represent false positives, potentially caused by conflicts within the \textsc{Slither} execution flow or interactions between its detectors.

\paragraph{Multiple calls in a loop (CL)} The CL operator introduces side effects that are strictly related to the vulnerability injected in 30\% of the cases. It introduces \textit{msg-value-loop} (6617), \textit{costly-loop} (451), and \textit{cyclomatic-complexity} (126). The first case occurs when \texttt{msg.value} is used in mathematical operations inside a loop. The second occurs when we use costly operations inside a loop that might waste gas, so optimizations are justified. The last case occurs when the cyclomatic complexity is higher than 10. In this case, all side effects are closely related to the mutation and, in some cases, are unavoidable.

\paragraph{Delegatecall to untrusted callee} The DTU operator introduces \textit{missing-zero-check} and \textit{naming-convention}. The mutation introduces a new address called \textit{delegate} and a new function to set the address called \textit{setDelegate}. The setter function does not check that the address is not the zero address (\textit{address(0)}). The \textit{naming-convention} is a warning that refers to the address argument of \textit{setDelegate}; it is not in \textit{mixedCase} violating the Style Guide of Solidity\footnote{https://docs.soliditylang.org/en/latest/}.

\examplebox{\faKey \hspace{0.2mm} \textbf{Discussion Summary.}
\textsc{Slither} exhibits inconsistent behavior and misses vulnerabilities in specific scenarios, suggesting potential gaps in its detection algorithms. Furthermore, its ability to detect some warnings depends on the presence of specific variables, even when these variables are not directly related to the core logic of the vulnerability.
}



\section{Threats To Validity}
\label{sec:ttv}
This section describes the potential threats to validity, including construct, internal, external, and conclusion validity.

\paragraph{Internal Validity}
The vulnerability injection approach threatens the validity of our study. The initial version of our mutation operators exhibits side effects, which can impact the results. As discussed in \Cref{sec:discussion}, some side effects are understandable and legitimate, others are unavoidable, and some can be mitigated by refining the mutation operators. 

Another threat concerns the relationship between the injected vulnerability and its side effects. Sometimes, the code smells that emerged as side effects were not attributable to the injected vulnerability. Mutations could introduce code smells independently of the injected vulnerability.  
We have discussed all observed side effects and explained their underlying causes, aiming to increase awareness about their proper usage and highlight areas for improvement in the mutation operators.

\paragraph{External Validity}
We relied on \textsc{Slither}~\cite{slither} as the static analysis tool to detect injected vulnerabilities, which may limit the generalizability of our findings regarding static analysis tools. However, \textsc{Slither} is one of the most widely used tools in smart contract analysis, recognized for its fast execution time and popularity in academic and industrial settings. It is open-source, actively maintained, and readily accessible, making it ideal for this study. Importantly, the experiment could be easily extended to evaluate additional tools.

Another external threat to validity is the focus on only six vulnerabilities. However, these vulnerabilities were selected because they are among the most frequently discussed in the literature~\cite{Zaazaa} and represent common security issues in smart contracts. In addition, the tool can be easily extended by implementing new mutation operators to inject new vulnerabilities.

\paragraph{Construct Validity}
A threat to construct validity concerns the analysis of false negatives. An exhaustive analysis of all false negatives would take too much time and might reveal cases we did not consider. We restricted the false negative analysis to observing a statistically significant sample from which we extracted the findings reported in the paper. 

A second threat concerns using \textsc{SmartBugs}~\cite{smartbugs_tool}, which accelerates the experiments and simplifies the analysis of the results. We followed its official documentation, which lists some issues and discusses how to mitigate them. Nevertheless, we acknowledge that our study could have been threatened by relying on this framework.

Another threat to construct validity relates to \muse{} implementation. Our tool is based upon \textsc{SuMo}~\cite{sumo}, a widely recognized and tested mutation testing tool that has already been extended for regression mutation testing~\cite{resumo}. However, by relying on \textsc{SuMo}, \muse{} may inherit its defects, potentially impacting the performance of \muse{}. To mitigate the potential issue, we manually validated our extensions using a statistically significant set of smart contracts composed of a random subset of the smartbugs-wild dataset.

Lastly, choosing the dataset for our experiments and validation introduces a potential threat to external and construct validity. We selected the smartbugs-wild dataset~\cite{smartbugs_wild}, one of the most widely used datasets in the literature. The dataset is recognized for its considerable size and frequent use in empirical investigations, reinforcing its relevance and reliability for our study.

\section{Conclusion}
\label{sec:conclusion}

This paper explores a mutation-based approach to inject known vulnerabilities into smart contracts to generate new benchmarks to evaluate vulnerability detection tools. We proposed \muse{}, a mutation seeding tool that implements mutational operators that identify the appropriate pattern in which to inject the mutation. \muse{} is capable of injecting six vulnerabilities. An injection rate characterizes each vulnerability since not all smart contracts have the conditions to be mutated and thus be vulnerable to a problem. \muse{} has been validated and is easily extended by adding new mutational operators. \textsc{Slither}, a static analyzer, analyzed smart contracts generated by mutation. The results showed gaps in \textsc{Slither} detectors, showing the current limitations of static analyzers and leaving room for improvement. Our study highlights that using new benchmarks generated through a mutation-based approach can improve the validation of static analysis tools. 

As part of our future work, we aim to enhance \muse{} by introducing additional mutation operators and refining existing ones to reduce unintended behavior. Our ultimate goal is to develop a fully automated tool capable of generating vulnerable smart contracts starting from an initial set of smart contracts. We plan to address the gaps identified by Bobadilla \textit{et al.}~\cite{automatedFixes} by creating new datasets. The current literature offers limited labeled benchmarks, most of which consist of smart contracts written in older versions of Solidity. Using our mutation seeding tool, we intend to inject vulnerabilities into audited and updated smart contracts to produce new, labeled benchmarks and fill these critical gaps.

\begin{acks}
  Finanziato dall’Unione Europea - Next Generation EU, Missione 4 Componente 1 CUP D53D23008400006.
\end{acks}

\bibliographystyle{ACM-Reference-Format}
\bibliography{main}


\begin{thebibliography}{40}


\ifx \showCODEN    \undefined \def \showCODEN     #1{\unskip}     \fi
\ifx \showISBNx    \undefined \def \showISBNx     #1{\unskip}     \fi
\ifx \showISBNxiii \undefined \def \showISBNxiii  #1{\unskip}     \fi
\ifx \showISSN     \undefined \def \showISSN      #1{\unskip}     \fi
\ifx \showLCCN     \undefined \def \showLCCN      #1{\unskip}     \fi
\ifx \shownote     \undefined \def \shownote      #1{#1}          \fi
\ifx \showarticletitle \undefined \def \showarticletitle #1{#1}   \fi
\ifx \showURL      \undefined \def \showURL       {\relax}        \fi
\providecommand\bibfield[2]{#2}
\providecommand\bibinfo[2]{#2}
\providecommand\natexlab[1]{#1}
\providecommand\showeprint[2][]{arXiv:#2}

\bibitem[Atzei et~al\mbox{.}(2017)]%
        {attacks}
\bibfield{author}{\bibinfo{person}{Nicola Atzei}, \bibinfo{person}{Massimo Bartoletti}, {and} \bibinfo{person}{Tiziana Cimoli}.} \bibinfo{year}{2017}\natexlab{}.
\newblock \showarticletitle{A survey of attacks on ethereum smart contracts (sok)}. In \bibinfo{booktitle}{\emph{Principles of Security and Trust: 6th International Conference, POST 2017, Held as Part of the European Joint Conferences on Theory and Practice of Software, ETAPS 2017, Uppsala, Sweden, April 22-29, 2017, Proceedings 6}}. Springer, \bibinfo{pages}{164--186}.
\newblock


\bibitem[Barboni et~al\mbox{.}(2021)]%
        {sumo}
\bibfield{author}{\bibinfo{person}{Morena Barboni}, \bibinfo{person}{Andrea Morichetta}, {and} \bibinfo{person}{Andrea Polini}.} \bibinfo{year}{2021}\natexlab{}.
\newblock \showarticletitle{SuMo: A Mutation Testing Strategy for Solidity Smart Contracts}. In \bibinfo{booktitle}{\emph{2021 IEEE/ACM International Conference on Automation of Software Test (AST)}}. \bibinfo{pages}{50--59}.
\newblock
\href{https://doi.org/10.1109/AST52587.2021.00014}{doi:\nolinkurl{10.1109/AST52587.2021.00014}}


\bibitem[Barboni et~al\mbox{.}(2022)]%
        {sumo_journal}
\bibfield{author}{\bibinfo{person}{Morena Barboni}, \bibinfo{person}{Andrea Morichetta}, {and} \bibinfo{person}{Andrea Polini}.} \bibinfo{year}{2022}\natexlab{}.
\newblock \showarticletitle{SuMo: A mutation testing approach and tool for the Ethereum blockchain}.
\newblock \bibinfo{journal}{\emph{Journal of Systems and Software}}  \bibinfo{volume}{193} (\bibinfo{year}{2022}), \bibinfo{pages}{111445}.
\newblock
\showISSN{0164-1212}
\href{https://doi.org/10.1016/j.jss.2022.111445}{doi:\nolinkurl{10.1016/j.jss.2022.111445}}


\bibitem[Barboni et~al\mbox{.}(2024)]%
        {resumo}
\bibfield{author}{\bibinfo{person}{Morena Barboni}, \bibinfo{person}{Andrea Morichetta}, \bibinfo{person}{Andrea Polini}, {and} \bibinfo{person}{Francesco Casoni}.} \bibinfo{year}{2024}\natexlab{}.
\newblock \showarticletitle{ReSuMo: a regression strategy and tool for mutation testing of solidity smart contracts}.
\newblock \bibinfo{journal}{\emph{Software Quality Journal}} \bibinfo{volume}{32}, \bibinfo{number}{1} (\bibinfo{year}{2024}), \bibinfo{pages}{225--253}.
\newblock


\bibitem[Beck(2022)]%
        {tdd}
\bibfield{author}{\bibinfo{person}{Kent Beck}.} \bibinfo{year}{2022}\natexlab{}.
\newblock \bibinfo{booktitle}{\emph{Test driven development: By example}}.
\newblock \bibinfo{publisher}{Addison-Wesley Professional}.
\newblock


\bibitem[Bobadilla et~al\mbox{.}(2025)]%
        {automatedFixes}
\bibfield{author}{\bibinfo{person}{Sofia Bobadilla}, \bibinfo{person}{Monica Jin}, {and} \bibinfo{person}{Martin Monperrus}.} \bibinfo{year}{2025}\natexlab{}.
\newblock \showarticletitle{Do Automated Fixes Truly Mitigate Smart Contract Exploits?}
\newblock \bibinfo{journal}{\emph{arXiv preprint arXiv:2501.04600}} (\bibinfo{year}{2025}).
\newblock


\bibitem[Buterin et~al\mbox{.}(2014)]%
        {buterin_ethereum}
\bibfield{author}{\bibinfo{person}{Vitalik Buterin} {et~al\mbox{.}}} \bibinfo{year}{2014}\natexlab{}.
\newblock \showarticletitle{Ethereum white paper: a next generation smart contract \& decentralized application platform}.
\newblock \bibinfo{journal}{\emph{First version}}  \bibinfo{volume}{53} (\bibinfo{year}{2014}).
\newblock


\bibitem[Chapman et~al\mbox{.}(2019)]%
        {Deviant}
\bibfield{author}{\bibinfo{person}{Patrick Chapman}, \bibinfo{person}{Dianxiang Xu}, \bibinfo{person}{Lin Deng}, {and} \bibinfo{person}{Yin Xiong}.} \bibinfo{year}{2019}\natexlab{}.
\newblock \showarticletitle{Deviant: A Mutation Testing Tool for Solidity Smart Contracts}. In \bibinfo{booktitle}{\emph{2019 IEEE International Conference on Blockchain (Blockchain)}}. \bibinfo{pages}{319--324}.
\newblock
\href{https://doi.org/10.1109/Blockchain.2019.00050}{doi:\nolinkurl{10.1109/Blockchain.2019.00050}}


\bibitem[Chen et~al\mbox{.}(2018)]%
        {understendingSC}
\bibfield{author}{\bibinfo{person}{Ting Chen}, \bibinfo{person}{Yuxiao Zhu}, \bibinfo{person}{Zihao Li}, \bibinfo{person}{Jiachi Chen}, \bibinfo{person}{Xiaoqi Li}, \bibinfo{person}{Xiapu Luo}, \bibinfo{person}{Xiaodong Lin}, {and} \bibinfo{person}{Xiaosong Zhange}.} \bibinfo{year}{2018}\natexlab{}.
\newblock \showarticletitle{Understanding Ethereum via Graph Analysis}. In \bibinfo{booktitle}{\emph{IEEE INFOCOM 2018 - IEEE Conference on Computer Communications}}. \bibinfo{pages}{1484--1492}.
\newblock
\href{https://doi.org/10.1109/INFOCOM.2018.8486401}{doi:\nolinkurl{10.1109/INFOCOM.2018.8486401}}


\bibitem[Chu et~al\mbox{.}(2024)]%
        {SGDL_rw}
\bibfield{author}{\bibinfo{person}{Hanting Chu}, \bibinfo{person}{Pengcheng Zhang}, \bibinfo{person}{Hai Dong}, \bibinfo{person}{Yan Xiao}, {and} \bibinfo{person}{Shunhui Ji}.} \bibinfo{year}{2024}\natexlab{}.
\newblock \showarticletitle{SGDL: Smart contract vulnerability generation via deep learning}.
\newblock \bibinfo{journal}{\emph{Journal of Software: Evolution and Process}} \bibinfo{volume}{36}, \bibinfo{number}{12} (\bibinfo{year}{2024}), \bibinfo{pages}{e2712}.
\newblock


\bibitem[Daspe et~al\mbox{.}(2024)]%
        {daspe}
\bibfield{author}{\bibinfo{person}{Etienne Daspe}, \bibinfo{person}{Mathis Durand}, \bibinfo{person}{Julien Hatin}, {and} \bibinfo{person}{Salma Bradai}.} \bibinfo{year}{2024}\natexlab{}.
\newblock \showarticletitle{Benchmarking Large Language Models for Ethereum Smart Contract Development}. \bibinfo{pages}{1--4}.
\newblock
\href{https://doi.org/10.1109/BRAINS63024.2024.10732686}{doi:\nolinkurl{10.1109/BRAINS63024.2024.10732686}}


\bibitem[di~Angelo et~al\mbox{.}(2023)]%
        {smartbugs_tool}
\bibfield{author}{\bibinfo{person}{Monika di Angelo}, \bibinfo{person}{Thomas Durieux}, \bibinfo{person}{Jo{\~a}o~F. Ferreira}, {and} \bibinfo{person}{Gernot Salzer}.} \bibinfo{year}{2023}\natexlab{}.
\newblock \showarticletitle{{SmartBugs} 2.0: An Execution Framework for Weakness Detection in Ethereum Smart Contracts}. In \bibinfo{booktitle}{\emph{38th IEEE/ACM International Conference on Automated Software Engineering (ASE)}}. \bibinfo{publisher}{IEEE Computer Society}, \bibinfo{pages}{2102--2105}.
\newblock
\href{https://doi.org/10.1109/ASE56229.2023.00060}{doi:\nolinkurl{10.1109/ASE56229.2023.00060}}


\bibitem[Di~Angelo and Salzer(2019)]%
        {benchmark}
\bibfield{author}{\bibinfo{person}{Monika Di~Angelo} {and} \bibinfo{person}{Gernot Salzer}.} \bibinfo{year}{2019}\natexlab{}.
\newblock \showarticletitle{A survey of tools for analyzing ethereum smart contracts}. In \bibinfo{booktitle}{\emph{2019 IEEE international conference on decentralized applications and infrastructures (DAPPCON)}}. IEEE, \bibinfo{pages}{69--78}.
\newblock


\bibitem[Dika and Nowostawski(2018)]%
        {8726833}
\bibfield{author}{\bibinfo{person}{Ardit Dika} {and} \bibinfo{person}{Mariusz Nowostawski}.} \bibinfo{year}{2018}\natexlab{}.
\newblock \showarticletitle{Security Vulnerabilities in Ethereum Smart Contracts}. In \bibinfo{booktitle}{\emph{2018 IEEE International Conference on Internet of Things (iThings) and IEEE Green Computing and Communications (GreenCom) and IEEE Cyber, Physical and Social Computing (CPSCom) and IEEE Smart Data (SmartData)}}. \bibinfo{pages}{955--962}.
\newblock
\href{https://doi.org/10.1109/Cybermatics_2018.2018.00182}{doi:\nolinkurl{10.1109/Cybermatics_2018.2018.00182}}


\bibitem[Durieux et~al\mbox{.}(2020)]%
        {smartbugs_wild}
\bibfield{author}{\bibinfo{person}{Thomas Durieux}, \bibinfo{person}{Jo{\~a}o~F Ferreira}, \bibinfo{person}{Rui Abreu}, {and} \bibinfo{person}{Pedro Cruz}.} \bibinfo{year}{2020}\natexlab{}.
\newblock \showarticletitle{Empirical review of automated analysis tools on 47,587 ethereum smart contracts}. In \bibinfo{booktitle}{\emph{Proceedings of the ACM/IEEE 42nd International conference on software engineering}}. \bibinfo{pages}{530--541}.
\newblock


\bibitem[Feist et~al\mbox{.}(2019)]%
        {slither}
\bibfield{author}{\bibinfo{person}{Josselin Feist}, \bibinfo{person}{Gustavo Grieco}, {and} \bibinfo{person}{Alex Groce}.} \bibinfo{year}{2019}\natexlab{}.
\newblock \showarticletitle{Slither: a static analysis framework for smart contracts}. In \bibinfo{booktitle}{\emph{2019 IEEE/ACM 2nd International Workshop on Emerging Trends in Software Engineering for Blockchain (WETSEB)}}. IEEE, \bibinfo{pages}{8--15}.
\newblock


\bibitem[Ghaleb and Pattabiraman(2020)]%
        {solidifi}
\bibfield{author}{\bibinfo{person}{Asem Ghaleb} {and} \bibinfo{person}{Karthik Pattabiraman}.} \bibinfo{year}{2020}\natexlab{}.
\newblock \showarticletitle{`}. In \bibinfo{booktitle}{\emph{Proceedings of the 29th ACM SIGSOFT International Symposium on Software Testing and Analysis}} (Virtual Event, USA) \emph{(\bibinfo{series}{ISSTA 2020})}. \bibinfo{publisher}{Association for Computing Machinery}, \bibinfo{address}{New York, NY, USA}, \bibinfo{pages}{415–427}.
\newblock
\showISBNx{9781450380089}
\href{https://doi.org/10.1145/3395363.3397385}{doi:\nolinkurl{10.1145/3395363.3397385}}


\bibitem[Hajdu et~al\mbox{.}(2020)]%
        {FaultInjection}
\bibfield{author}{\bibinfo{person}{Ákos Hajdu}, \bibinfo{person}{Naghmeh Ivaki}, \bibinfo{person}{Imre Kocsis}, \bibinfo{person}{Attila Klenik}, \bibinfo{person}{László Gönczy}, \bibinfo{person}{Nuno Laranjeiro}, \bibinfo{person}{Henrique Madeira}, {and} \bibinfo{person}{András Pataricza}.} \bibinfo{year}{2020}\natexlab{}.
\newblock \showarticletitle{Using Fault Injection to Assess Blockchain Systems in Presence of Faulty Smart Contracts}.
\newblock \bibinfo{journal}{\emph{IEEE Access}}  \bibinfo{volume}{8} (\bibinfo{year}{2020}), \bibinfo{pages}{190760--190783}.
\newblock
\href{https://doi.org/10.1109/ACCESS.2020.3032239}{doi:\nolinkurl{10.1109/ACCESS.2020.3032239}}


\bibitem[Iuliano and Nucci(2024)]%
        {iulianoSLR}
\bibfield{author}{\bibinfo{person}{Gerardo Iuliano} {and} \bibinfo{person}{Dario~Di Nucci}.} \bibinfo{year}{2024}\natexlab{}.
\newblock \bibinfo{title}{Smart Contract Vulnerabilities, Tools, and Benchmarks: An Updated Systematic Literature Review}.
\newblock
\showeprint[arxiv]{2412.01719}~[cs.SE]
\urldef\tempurl%
\url{https://arxiv.org/abs/2412.01719}
\showURL{%
\tempurl}


\bibitem[Ivanova and Khritankov(2020)]%
        {RegularMutator}
\bibfield{author}{\bibinfo{person}{Y. Ivanova} {and} \bibinfo{person}{A. Khritankov}.} \bibinfo{year}{2020}\natexlab{}.
\newblock \showarticletitle{RegularMutator: A Mutation Testing Tool for Solidity Smart Contracts}.
\newblock \bibinfo{journal}{\emph{Procedia Computer Science}}  \bibinfo{volume}{178} (\bibinfo{year}{2020}), \bibinfo{pages}{75--83}.
\newblock
\showISSN{1877-0509}
\href{https://doi.org/10.1016/j.procs.2020.11.009}{doi:\nolinkurl{10.1016/j.procs.2020.11.009}}
\newblock
\shownote{9th International Young Scientists Conference in Computational Science, YSC2020, 05-12 September 2020}.


\bibitem[Jia and Harman(2011)]%
        {mutationTesting}
\bibfield{author}{\bibinfo{person}{Yue Jia} {and} \bibinfo{person}{Mark Harman}.} \bibinfo{year}{2011}\natexlab{}.
\newblock \showarticletitle{An Analysis and Survey of the Development of Mutation Testing}.
\newblock \bibinfo{journal}{\emph{IEEE Trans. Softw. Eng.}} \bibinfo{volume}{37}, \bibinfo{number}{5} (\bibinfo{date}{Sept.} \bibinfo{year}{2011}), \bibinfo{pages}{649–678}.
\newblock
\showISSN{0098-5589}
\href{https://doi.org/10.1109/TSE.2010.62}{doi:\nolinkurl{10.1109/TSE.2010.62}}


\bibitem[Jiang et~al\mbox{.}(2018)]%
        {contractfuzzer}
\bibfield{author}{\bibinfo{person}{Bo Jiang}, \bibinfo{person}{Ye Liu}, {and} \bibinfo{person}{Wing~Kwong Chan}.} \bibinfo{year}{2018}\natexlab{}.
\newblock \showarticletitle{Contractfuzzer: Fuzzing smart contracts for vulnerability detection}. In \bibinfo{booktitle}{\emph{Proceedings of the 33rd ACM/IEEE international conference on automated software engineering}}. \bibinfo{pages}{259--269}.
\newblock


\bibitem[Kushwaha et~al\mbox{.}(2022a)]%
        {9667515}
\bibfield{author}{\bibinfo{person}{Satpal~Singh Kushwaha}, \bibinfo{person}{Sandeep Joshi}, \bibinfo{person}{Dilbag Singh}, \bibinfo{person}{Manjit Kaur}, {and} \bibinfo{person}{Heung-No Lee}.} \bibinfo{year}{2022}\natexlab{a}.
\newblock \showarticletitle{Systematic Review of Security Vulnerabilities in Ethereum Blockchain Smart Contract}.
\newblock \bibinfo{journal}{\emph{IEEE Access}}  \bibinfo{volume}{10} (\bibinfo{year}{2022}), \bibinfo{pages}{6605--6621}.
\newblock
\href{https://doi.org/10.1109/ACCESS.2021.3140091}{doi:\nolinkurl{10.1109/ACCESS.2021.3140091}}


\bibitem[Kushwaha et~al\mbox{.}(2022b)]%
        {kushwaha}
\bibfield{author}{\bibinfo{person}{Satpal~Singh Kushwaha}, \bibinfo{person}{Sandeep Joshi}, \bibinfo{person}{Dilbag Singh}, \bibinfo{person}{Manjit Kaur}, {and} \bibinfo{person}{Heung-No Lee}.} \bibinfo{year}{2022}\natexlab{b}.
\newblock \showarticletitle{Systematic review of security vulnerabilities in ethereum blockchain smart contract}.
\newblock \bibinfo{journal}{\emph{IEEE Access}}  \bibinfo{volume}{10} (\bibinfo{year}{2022}), \bibinfo{pages}{6605--6621}.
\newblock


\bibitem[Liu et~al\mbox{.}(2018)]%
        {reguard}
\bibfield{author}{\bibinfo{person}{Chao Liu}, \bibinfo{person}{Han Liu}, \bibinfo{person}{Zhao Cao}, \bibinfo{person}{Zhong Chen}, \bibinfo{person}{Bangdao Chen}, {and} \bibinfo{person}{Bill Roscoe}.} \bibinfo{year}{2018}\natexlab{}.
\newblock \showarticletitle{Reguard: finding reentrancy bugs in smart contracts}. In \bibinfo{booktitle}{\emph{Proceedings of the 40th International Conference on Software Engineering: Companion Proceeedings}}. \bibinfo{pages}{65--68}.
\newblock


\bibitem[Luu et~al\mbox{.}(2016)]%
        {smarter_sc}
\bibfield{author}{\bibinfo{person}{Loi Luu}, \bibinfo{person}{Duc-Hiep Chu}, \bibinfo{person}{Hrishi Olickel}, \bibinfo{person}{Prateek Saxena}, {and} \bibinfo{person}{Aquinas Hobor}.} \bibinfo{year}{2016}\natexlab{}.
\newblock \showarticletitle{Making smart contracts smarter}. In \bibinfo{booktitle}{\emph{Proceedings of the 2016 ACM SIGSAC conference on computer and communications security}}. \bibinfo{pages}{254--269}.
\newblock


\bibitem[Mehar et~al\mbox{.}(2019)]%
        {dao_attack}
\bibfield{author}{\bibinfo{person}{Muhammad~Izhar Mehar}, \bibinfo{person}{Charles~Louis Shier}, \bibinfo{person}{Alana Giambattista}, \bibinfo{person}{Elgar Gong}, \bibinfo{person}{Gabrielle Fletcher}, \bibinfo{person}{Ryan Sanayhie}, \bibinfo{person}{Henry~M Kim}, {and} \bibinfo{person}{Marek Laskowski}.} \bibinfo{year}{2019}\natexlab{}.
\newblock \showarticletitle{Understanding a revolutionary and flawed grand experiment in blockchain: the DAO attack}.
\newblock \bibinfo{journal}{\emph{Journal of Cases on Information Technology (JCIT)}} \bibinfo{volume}{21}, \bibinfo{number}{1} (\bibinfo{year}{2019}), \bibinfo{pages}{19--32}.
\newblock


\bibitem[Mense and Flatscher(2018)]%
        {mense}
\bibfield{author}{\bibinfo{person}{Alexander Mense} {and} \bibinfo{person}{Markus Flatscher}.} \bibinfo{year}{2018}\natexlab{}.
\newblock \showarticletitle{Security vulnerabilities in ethereum smart contracts}. In \bibinfo{booktitle}{\emph{Proceedings of the 20th international conference on information integration and web-based applications \& services}}. \bibinfo{pages}{375--380}.
\newblock


\bibitem[Papadakis et~al\mbox{.}(2019)]%
        {PAPADAKIS}
\bibfield{author}{\bibinfo{person}{Mike Papadakis}, \bibinfo{person}{Marinos Kintis}, \bibinfo{person}{Jie Zhang}, \bibinfo{person}{Yue Jia}, \bibinfo{person}{Yves~Le Traon}, {and} \bibinfo{person}{Mark Harman}.} \bibinfo{year}{2019}\natexlab{}.
\newblock \showarticletitle{Chapter Six - Mutation Testing Advances: An Analysis and Survey}.
\newblock \bibinfo{series}{Advances in Computers}, Vol.~\bibinfo{volume}{112}. \bibinfo{publisher}{Elsevier}, \bibinfo{pages}{275--378}.
\newblock
\showISSN{0065-2458}
\href{https://doi.org/10.1016/bs.adcom.2018.03.015}{doi:\nolinkurl{10.1016/bs.adcom.2018.03.015}}


\bibitem[Parizi et~al\mbox{.}(2018)]%
        {fp2}
\bibfield{author}{\bibinfo{person}{Reza~M Parizi}, \bibinfo{person}{Ali Dehghantanha}, \bibinfo{person}{Kim-Kwang~Raymond Choo}, {and} \bibinfo{person}{Amritraj Singh}.} \bibinfo{year}{2018}\natexlab{}.
\newblock \showarticletitle{Empirical vulnerability analysis of automated smart contracts security testing on blockchains}.
\newblock \bibinfo{journal}{\emph{arXiv preprint arXiv:1809.02702}} (\bibinfo{year}{2018}).
\newblock


\bibitem[Rameder et~al\mbox{.}(2022)]%
        {ramederSLR}
\bibfield{author}{\bibinfo{person}{Heidelinde Rameder}, \bibinfo{person}{Monika Di~Angelo}, {and} \bibinfo{person}{Gernot Salzer}.} \bibinfo{year}{2022}\natexlab{}.
\newblock \showarticletitle{Review of automated vulnerability analysis of smart contracts on Ethereum}.
\newblock \bibinfo{journal}{\emph{Frontiers in Blockchain}}  \bibinfo{volume}{5} (\bibinfo{year}{2022}), \bibinfo{pages}{814977}.
\newblock


\bibitem[Ren et~al\mbox{.}(2021)]%
        {ren}
\bibfield{author}{\bibinfo{person}{Meng Ren}, \bibinfo{person}{Zijing Yin}, \bibinfo{person}{Fuchen Ma}, \bibinfo{person}{Zhenyang Xu}, \bibinfo{person}{Yu Jiang}, \bibinfo{person}{Chengnian Sun}, \bibinfo{person}{Huizhong Li}, {and} \bibinfo{person}{Yan Cai}.} \bibinfo{year}{2021}\natexlab{}.
\newblock \showarticletitle{Empirical evaluation of smart contract testing: What is the best choice?}. In \bibinfo{booktitle}{\emph{Proceedings of the 30th ACM SIGSOFT international symposium on software testing and analysis}}. \bibinfo{pages}{566--579}.
\newblock


\bibitem[Samreen and Alalfi(2021)]%
        {smartscan}
\bibfield{author}{\bibinfo{person}{Noama~Fatima Samreen} {and} \bibinfo{person}{Manar~H Alalfi}.} \bibinfo{year}{2021}\natexlab{}.
\newblock \showarticletitle{Smartscan: an approach to detect denial of service vulnerability in ethereum smart contracts}. In \bibinfo{booktitle}{\emph{2021 IEEE/ACM 4th International Workshop on Emerging Trends in Software Engineering for Blockchain (WETSEB)}}. IEEE, \bibinfo{pages}{17--26}.
\newblock


\bibitem[Sayeed et~al\mbox{.}(2020)]%
        {sayeed}
\bibfield{author}{\bibinfo{person}{Sarwar Sayeed}, \bibinfo{person}{Hector Marco-Gisbert}, {and} \bibinfo{person}{Tom Caira}.} \bibinfo{year}{2020}\natexlab{}.
\newblock \showarticletitle{Smart contract: Attacks and protections}.
\newblock \bibinfo{journal}{\emph{Ieee Access}}  \bibinfo{volume}{8} (\bibinfo{year}{2020}), \bibinfo{pages}{24416--24427}.
\newblock


\bibitem[Sch{\"a}r(2021)]%
        {de-fi}
\bibfield{author}{\bibinfo{person}{Fabian Sch{\"a}r}.} \bibinfo{year}{2021}\natexlab{}.
\newblock \showarticletitle{Decentralized finance: On blockchain-and smart contract-based financial markets}.
\newblock \bibinfo{journal}{\emph{FRB of St. Louis Review}} (\bibinfo{year}{2021}).
\newblock


\bibitem[Vidal et~al\mbox{.}(2024)]%
        {openscv}
\bibfield{author}{\bibinfo{person}{Fernando~Richter Vidal}, \bibinfo{person}{Naghmeh Ivaki}, {and} \bibinfo{person}{Nuno Laranjeiro}.} \bibinfo{year}{2024}\natexlab{}.
\newblock \showarticletitle{OpenSCV: an open hierarchical taxonomy for smart contract vulnerabilities}.
\newblock \bibinfo{journal}{\emph{Empirical Software Engineering}} \bibinfo{volume}{29}, \bibinfo{number}{4} (\bibinfo{year}{2024}), \bibinfo{pages}{101}.
\newblock


\bibitem[Wang et~al\mbox{.}(2022)]%
        {etherfuzz}
\bibfield{author}{\bibinfo{person}{Xiaoyin Wang}, \bibinfo{person}{Jiaze Sun}, \bibinfo{person}{Chunyang Hu}, \bibinfo{person}{Panpan Yu}, \bibinfo{person}{Bin Zhang}, {and} \bibinfo{person}{Donghai Hou}.} \bibinfo{year}{2022}\natexlab{}.
\newblock \showarticletitle{EtherFuzz: mutation fuzzing smart contracts for TOD vulnerability detection}.
\newblock \bibinfo{journal}{\emph{Wireless Communications and Mobile Computing}} \bibinfo{volume}{2022}, \bibinfo{number}{1} (\bibinfo{year}{2022}), \bibinfo{pages}{1565007}.
\newblock


\bibitem[Zaazaa and El~Bakkali(2023)]%
        {Zaazaa}
\bibfield{author}{\bibinfo{person}{Oualid Zaazaa} {and} \bibinfo{person}{Hanan El~Bakkali}.} \bibinfo{year}{2023}\natexlab{}.
\newblock \showarticletitle{A systematic literature review of undiscovered vulnerabilities and tools in smart contract technology}.
\newblock \bibinfo{journal}{\emph{Journal of Intelligent Systems}}  \bibinfo{volume}{32} (\bibinfo{date}{09} \bibinfo{year}{2023}).
\newblock
\href{https://doi.org/10.1515/jisys-2023-0038}{doi:\nolinkurl{10.1515/jisys-2023-0038}}


\bibitem[Zheng et~al\mbox{.}(2024)]%
        {TSE}
\bibfield{author}{\bibinfo{person}{Zibin Zheng}, \bibinfo{person}{Jianzhong Su}, \bibinfo{person}{Jiachi Chen}, \bibinfo{person}{David Lo}, \bibinfo{person}{Zhijie Zhong}, {and} \bibinfo{person}{Mingxi Ye}.} \bibinfo{year}{2024}\natexlab{}.
\newblock \showarticletitle{DAppSCAN: Building Large-Scale Datasets for Smart Contract Weaknesses in DApp Projects}.
\newblock \bibinfo{journal}{\emph{IEEE Trans. Softw. Eng.}} \bibinfo{volume}{50}, \bibinfo{number}{6} (\bibinfo{date}{June} \bibinfo{year}{2024}), \bibinfo{pages}{1360–1373}.
\newblock
\showISSN{0098-5589}
\href{https://doi.org/10.1109/TSE.2024.3383422}{doi:\nolinkurl{10.1109/TSE.2024.3383422}}


\bibitem[Zou et~al\mbox{.}(2021)]%
        {8847638}
\bibfield{author}{\bibinfo{person}{Weiqin Zou}, \bibinfo{person}{David Lo}, \bibinfo{person}{Pavneet~Singh Kochhar}, \bibinfo{person}{Xuan-Bach~Dinh Le}, \bibinfo{person}{Xin Xia}, \bibinfo{person}{Yang Feng}, \bibinfo{person}{Zhenyu Chen}, {and} \bibinfo{person}{Baowen Xu}.} \bibinfo{year}{2021}\natexlab{}.
\newblock \showarticletitle{Smart Contract Development: Challenges and Opportunities}.
\newblock \bibinfo{journal}{\emph{IEEE Transactions on Software Engineering}} \bibinfo{volume}{47}, \bibinfo{number}{10} (\bibinfo{year}{2021}), \bibinfo{pages}{2084--2106}.
\newblock
\href{https://doi.org/10.1109/TSE.2019.2942301}{doi:\nolinkurl{10.1109/TSE.2019.2942301}}


\end{thebibliography}

\end{document}